\def\aa{{A\&A}}
\def\aas{{A\&AS}}

\def\apj{{ApJ}}
\def\apjs{{ApJS}}

\def\mnras{{MNRAS}}

\documentclass[cup5b]{caps}
\usepackage{graphicx}
\usepackage{amssymb}
\usepackage{ociwsymp4}   

\HeadText{F. Matteucci}  

\def\plotone#1{\centering \leavevmode
\includegraphics[width=.95\columnwidth]{#1}}

\def\plottwo#1#2{\centering \leavevmode
\includegraphics[width=.45\columnwidth]{#1} \hfil
\includegraphics[width=.45\columnwidth]{#2}}

\def\plotone#1{\centering \leavevmode
\includegraphics[width=.95\columnwidth]{#1}}

\def\plottwo#1#2{\centering \leavevmode
\includegraphics[width=.45\columnwidth]{#1} \hfil
\includegraphics[width=.45\columnwidth]{#2}}

\begin{document}

\pagenumbering{arabic}

\author[]{FRANCESCA MATTEUCCI\\University of Trieste, Astronomy Department, Via G.B.
Tiepolo 11, 34100 Trieste, Italy}

%
%

\chapter{Models of Chemical Evolution}

\begin{abstract}
The basic principles underlying galactic chemical evolution and the most important results of chemical evolution models are discussed. In particular, the chemical evolution of the Milky Way  galaxy, for which we possess
the majority of observational constraints, is described. Then, it is shown
how different star formation histories influence the chemical evolution of galaxies of different morphological type. Finally, the role of abundances and abundance ratios as cosmic clocks is emphasized and a comparison between model predictions and abundance patterns in high redshift objects is used to infer the nature and the age of these systems.

\end{abstract}

\section{Introduction}
In order to build a chemical evolution model one needs to specify some basic parameters such as the boundary conditions, namely whether the system is closed or open and whether the gas is primordial or already chemically enriched. Then the stellar birthate function, which is generally expressed as the product of two independent functions, the star formation rate (SFR)  and the initial mass function (IMF), namely:
\begin{equation}
B(m,t)=\psi(t) \varphi(m)
\end{equation}
where the SFR is assumed to be only a function of time and the IMF only a function of the stellar mass.
The stellar evolution and nucleosynthesis are also  necessary ingredients for modelling chemical evolution and in particular we need to specify the stellar yields ($p_{i}(m)$).

Then, one can include in the chemical evolution model some supplementary ingredients such as the the infall of extragalactic material, radial flows and galactic winds, which can play a more or less important role depending on the galactic system under study. In particular, gas infall and radial flows can be  important in describing the chemical evolution of spiral disks, whereas galactic outflows are probably important in elliptical galaxies where the SFR was very 
high in the past and there is no cold gas at the present time,  as well as in dwarf galaxies which possess a smaller potential well.
In fact, galactic outflows, which eventually can transform into galactic winds, are determined by two main processes; the supernova feedback, namely how much energy is transferred from SNe into the interstellar medium (ISM), and the galactic potential well.

In this paper, I will recall the most popular approximations used to describe both the basic and the supplementary ingredients involved in galactic chemical evolution.
Then I will describe detailed numerical models which can follow the evolution in space and time of the abundances of the most abundant chemical species. I will show some applications of such models to the Milky Way galaxy and then to galaxies of different morphological type, focussing on the fact
that abundances and abundance ratios are very useful tools to infer the history of star formation in galaxies, to impose constraints on the stellar nucleosynthesis and to derive information on the nature and the age of high redshift objects.

\section{The star formation rate}
The SFR is one of the most important drivers of galactic chemical evolution:
it describes the rate at which the gas is turned into stars in galaxies.
Since the physics of the star formation process is still not well known, several parametrizations are used to describe the SFR. A common aspect to the different formulations of the SFR is that they include a dependence upon the gas density.
Here I recall the most commonly used  parametrizations for the SFR adopted so far in the literature.

An exponentially decreasing SFR provides an easy to handle formula:
\begin{equation}
SFR= \nu e^{-t/ \tau_*}
\end{equation}
with $\tau_* = 5-15$ Gyr  in order to obtain a good fit to the properties of the solar neighbourhhood (Tosi, 1988) and $\nu=1-2 Gyr^{-1}$,  being the so-called efficiency of star formation 
which is expressed as the inverse of the timescale of star formation.

However, the most famous formulation and most widely adopted for the SFR is the Schmidt (1955) law:
\begin{equation}
SFR = \nu \sigma_{gas}^{k}
\end{equation}
which assumes that the SFR is proportional to some power of the volume or surface gas density. 
The exponent suggested by Schmidt was $k=2$ but Kennicutt (1998) suggested that
the best fit to the  observational data on spiral disks and starburst galaxies is obtained with an exponent $k=1.4 \pm 0.15$.

A more complex formulation, including a dependence also from the total surface mass density, which is induced by the SN feedback, was suggested by the observations of
Dopita \& Ryder (1994) who proposed the following formulation:

\begin{equation}
SFR= \nu \sigma_{tot}^{k_1} \sigma_{gas}^{k_2}
\end{equation}
with $1.5 < k_1 + k_2 < 2.5$.

Kennicutt suggested also an alternative law to the Schmidt-like one discussed above, in particular a law containing the angular rotation speed of gas, $\Omega_{gas}$:
\begin{equation}
SFR= 0.017 \Omega_{gas} \sigma_{gas}\propto R^{-1} \sigma_{gas}
\end{equation}

A similar law for the SFR taking into account star formation  induced by spiral density waves
was proposed by
Wyse \& Silk (1989) and it can be expressed as (Prantzos 2002):
\begin{equation}
SFR= \nu V(R)R^{-1} \sigma_{gas}^{1.5}
\end{equation}
where $V(R)$ is the rotational velocity in the disk and R is the galactocentric distance. 
It is worth noting that the SFRs expressed by eqs. (1.4), (1.5) and (1.6) contain a stronger dependence on the radial properties of the disk than the simple Schmidt law, and this characteristic is required to best fit the disk properties (see next sections).

\section{The IMF}

The most common parametrization for the IMF is that proposed by Salpeter (1955), which assumes a
one-slope power law  with $x=1.35$, in particular:

\begin{equation}
\varphi(M)=cM^{-(1+x)}
\end{equation}
is the number of stars with masses in the interval M, M+dM, and $c$ is a normalization constant.

The IMF is generally normalized as:
\begin{equation}
\int^{\infty}_{0}{M\varphi(M)dM}=1
\end{equation}

More recently, multi-slope ($x_1$, $x_2$,...)
expressions of the IMF
have been adopted since they are better describing the luminosity function of the main sequence stars in the solar vicinity
(Scalo 1986, 1998; Kroupa et al. 1993).
Generally, the IMF is assumed to be constant in space and time, with some exceptions such as the one suggested by Larson (1998), which adopts a variable slope:
\begin{equation}
x=1.35(1 + m/m_1)^{-1}
\end{equation}
where $m_1$ is variable with time and associated to the Jeans mass.
The effects of a variable IMF on the galactic disk properties have been studied by Chiappini et al. (2000),
who concluded that only a very ``ad hoc'' variation of the IMF can reproduce the majority of observational constraints,
thus favoring chemical evolution models with IMF constant in space and time.

\section{Infall and outflow}
Depending on the galactic system, the infall rate can be assumed to be 
constant in space and time, or more realistically
the infall rate can be variable in space and time:
\begin{equation}
IR= A(R) e^{-t/ \tau(R)}
\end{equation}
with $\tau(R)$ constant or varying along the disk.
The parameter $A(R)$ is derived by fitting the present time total surface mass density in the disk of the Galaxy, $\sigma_{tot}(t_G)$. 
Otherwise, for the formation of the Galaxy one can assume two 
independent episodes of infall during which the halo and perhaps part of the thick-disk formed first and then the thin-disk, respectively:
\begin{equation}
IR= A(R) e^{-t/ \tau_{H}(R)}+ B(R) e^{-(t-t_{max})/ \tau_{D}(R)}
\end{equation}
as in the two-infall model of Chiappini et al. (1997). Here $\tau_H$ is the timescale for the formation of the halo/thick disk and
$\tau_D(R)$ is the timescale for the formation of the thin-disk. This latter is assumed to increase linearly with galactocentric distance (Matteucci \& Fran\c cois 1989; Chiappini et al. 1997; Boissier \& Prantzos 1999).
For the rate of gas outflow or galactic wind there are no specific prescriptions but generally one simply assumes that the wind rate is proportional to the star formation rate through a suitable parameter (Hartwick, 1976; Matteucci \& Chiosi, 1983): 
\begin{equation}
W=- \lambda SFR
\end{equation}
with $\lambda$ being a free parameter.

\section{Stellar Yields}
The stellar yields are fundamental ingredients in galactic chemical evolution. In order to introduce the stellar yields we recall some useful concepts and define the {\it yield per stellar generation}.

Under the assumption of Instantaneous Recycling Approximation (I.R.A.)
we define the yield per stellar generation of a given element $i$ as (Tinsley, 1980): 
\begin{equation}
y_{i}={1 \over 1-R} \int^{\infty}_{1}{m p_{im} \varphi(m) dm} 
\end{equation}
where $p_{im}$ is the stellar yield of the element $i$, namely the newly 
formed
and ejected mass fraction of the element $i$ by a star of initial mass $m$.
It is worth noting that the expression (1.13) is an oversemplification and it does not have much 
meaning when considering chemical elements formed on long timescales.
The quantity $R$ is the returned fraction:
\begin{equation}
R=\int^{\infty}_{1}{(m-M_{rem}) \varphi(m) dm} 
\end{equation}
where $M_{rem}$ is the remnant mass of a star of mass m.
In the past ten years a large number of calculations
of the stellar yields, $p_{im}$, has 
become available
for stars of all masses and metallicities.
However, uncertainties in stellar yields are still present especially 
in the yields of Fe-peak elements. This is due to
uncertainties in the nuclear reaction rates, treatment of convection, mass cut, explosion energies, neutron fluxes and possible fall-back of matter on the proto-neutron star. Moreover, the $^{14}$N nucleosynthesis and its primary and/or secondary nature are still under debate. 
The most recent calculations are summarized below:
\begin{itemize}
\item Low and Intermediate mass stars ($0.8 \le M/M_{\odot} \le 8.0$)
(Marigo et al. 1996; van den Hoeck \& Groenewegen 1997;
Meynet \& Maeder, 2002;  Ventura et al. 2002;
Siess et al. 2002). These stars  produce $^{4}He$, C, N and some s-process ($A>90$) elements.

\item Massive stars ($M  \ge 10 M_{\odot}$)
(Woosley \& Weaver 1995; 
Langer \& Henkel 1995;  Thielemann et al. 1996; Nomoto et al. 1997;
Rauscher et al. 2002; Limongi \& Chieffi 2003). These stars  produce  mainly
$\alpha$-elements (O, Ne, Mg, Si, S, Ca), some Fe-peak elements,  
s-process elements ($A< 90$) and r-process elements.

\item Type Ia SNe (Nomoto et al. 1997;
Iwamoto et al. 1999) produce mainly Fe-peak elements.

\item Very massive objects ($M> 100 M_{\odot}$), if they exist, should produce mostly oxygen 
(Portinari et al. 1998; Umeda \& Nomoto 2001; Nakamura et al. 2001). 

\end{itemize}
In Figures 1.1 and  1.2 we show a comparison between the yields, produced in massive stars, of two $\alpha$-elements (O and Mg) and Fe, respectively, as
predicted by different authors. As is evident from Fig. 1.1 the O yields from different sources seem to be in good agreement with 
one another.
The yields of Mg, especially the most recent ones seem to be also in good agreement with each others, whereas for the yields of Fe (Figure 1.2)
there is not yet an agreement among different authors. Observational estimates of Fe produced in type II SNe (Elmhamdi et al. 2003) can help in constraining the Fe yields in massive stars.

\begin{figure}
\centering
\plottwo{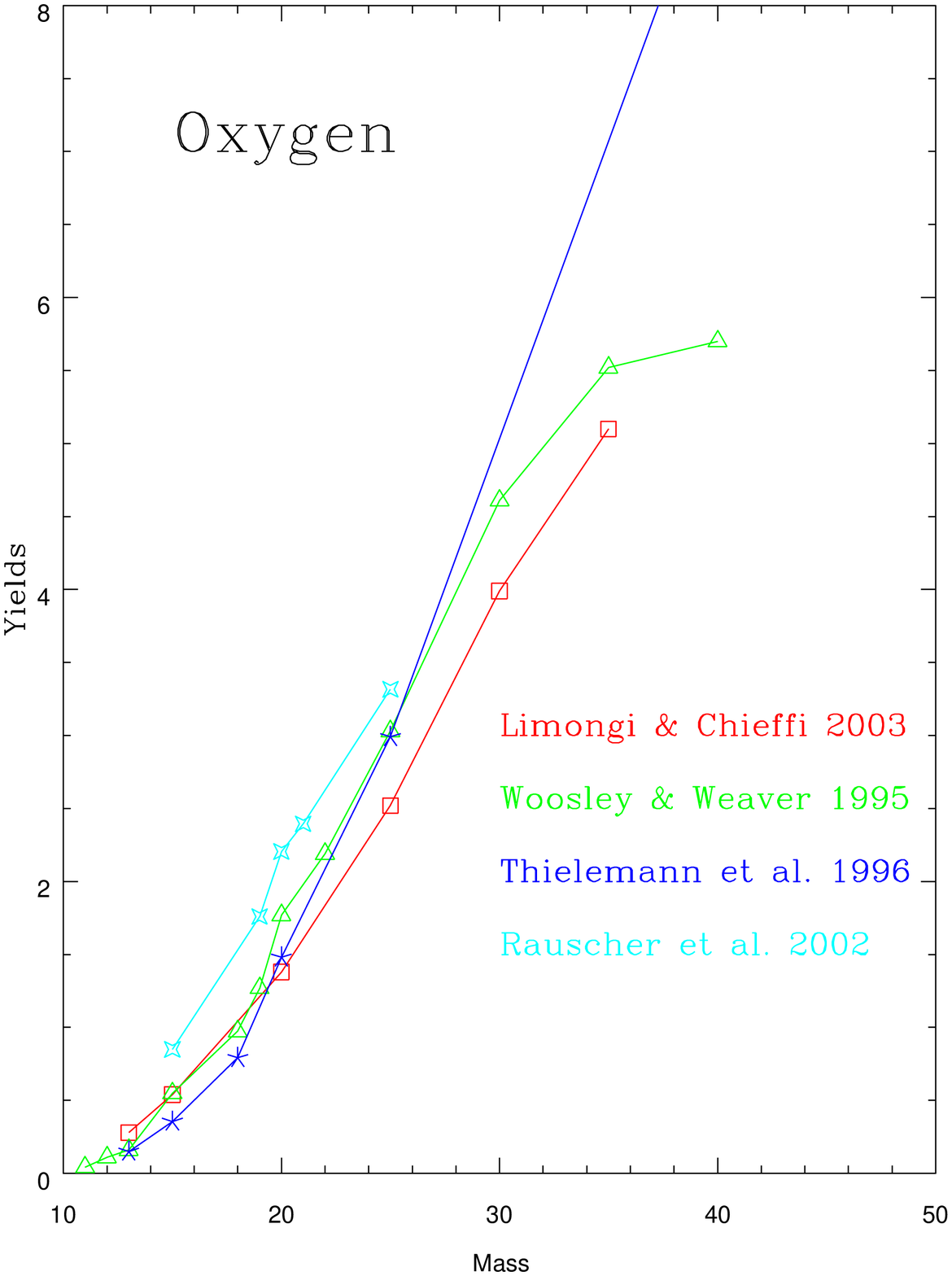}{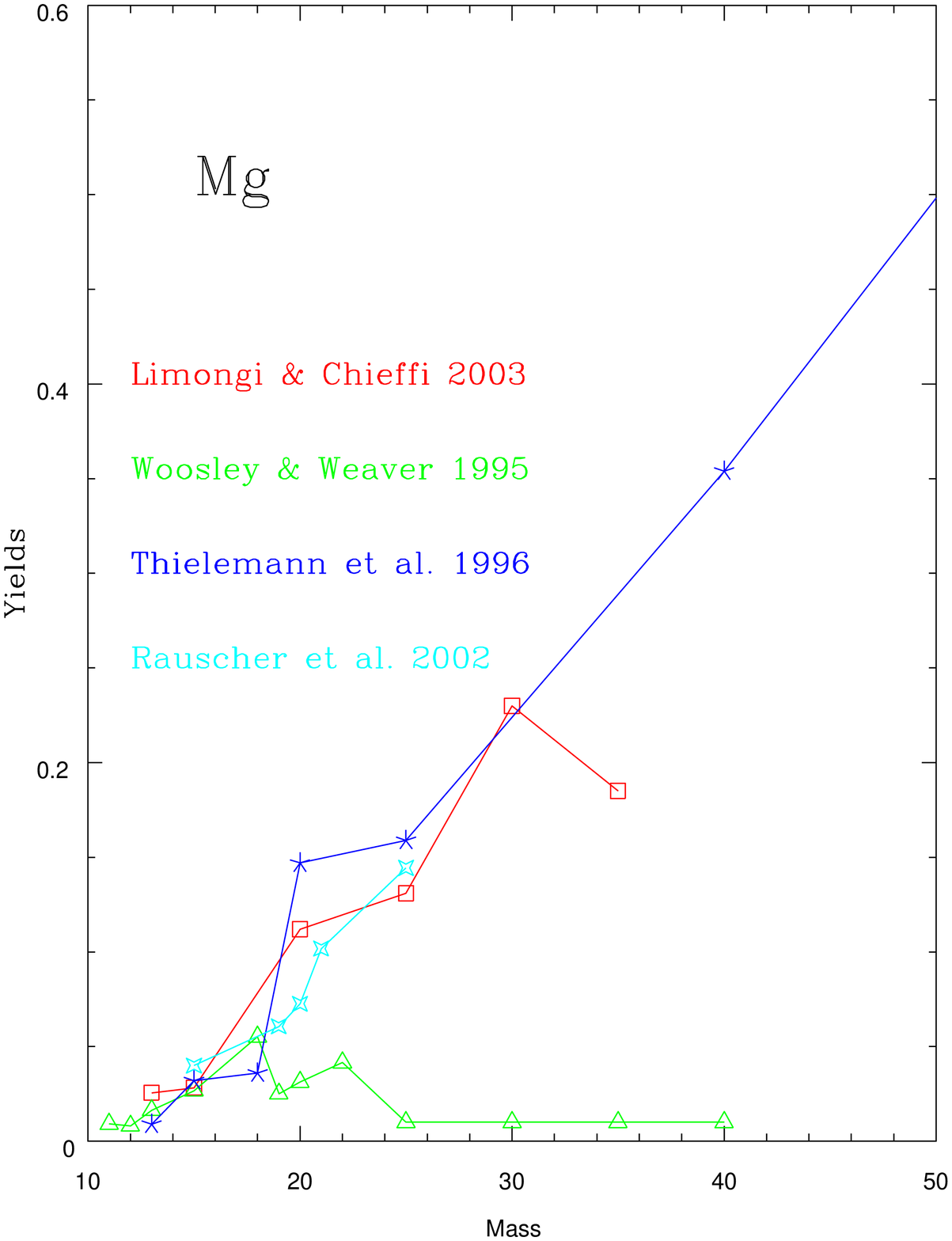}
\caption{Comparison between different yields of oxygen and Mg from  SNII: open triangles, Woosley \& Weaver (1995); open squares, Limongi \& Chieffi (2003); stars, Thielemann et al. (1996); four-point stars, Rauscher et al. (2002).
}
\end{figure}

\begin{figure}
\centering
\includegraphics[width=6cm,angle=0]{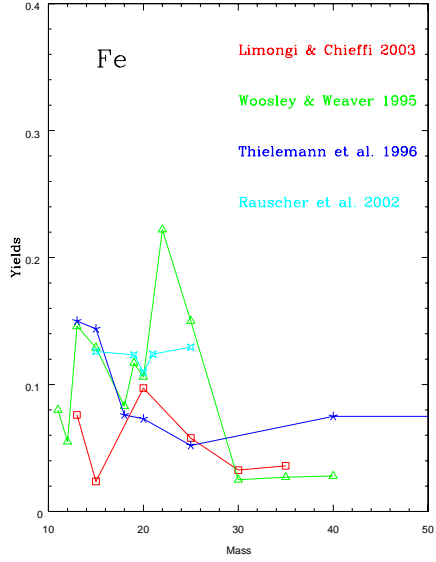}
\caption{Comparison between different yields of iron from  SNII: open triangles, Woosley\& Weaver 
(1995); open squares Limongi \& Chieffi (2003); stars Thielemann et al. (1996); four-point stars Rauscher et al. (2002).}
\end{figure}

\section{Analytical models of chemical evolution}

The simplest model of galactic chemical evolution is the so-called 
{\it Simple Model} for
the evolution of the solar neighbourhood. We define the solar neighbourhood or solar vicinity as a cylinder centered in the Sun with 1 kpc radius.

The basic assumptions of the Simple Model can be summarized as follows:
\begin{itemize}
\item the system is one-zone and closed, no inflows 
or outflows
\item the initial gas is primordial (no metals)
\item the instantaneous recycling approximation holds (I.R.A.) 
\item $\varphi(m)$ is constant in time and space
\item the gas is well mixed at any time (I.M.A.)
\end{itemize}

Let $X_i$ be the abundance of an element $i$ and 
$\beta= {M_{gas} \over M_{tot}}$ the ratio between the mass of gas and the total mass of the system.
If
$X_i<<1$, which is generally true for metals, then we can write:

\begin{equation}
X_i= y_{i} ln({ 1 \over \beta}) 
\end{equation}
which is the well known  solution for the Simple Model, where 
$y_i$ is the yield per stellar generation as defined in eq. (1.13). In particular,
the yield appearing in eq. (1.15) is usually referred to as the {\it effective yield}.
If $X_i$ is not much lower than 1, 
as is the case for $X_{He}$ (Maeder 1992), 
a more precise expression for the solution of the Simple Model is given by:
\begin{equation}
X_i=1- \beta^{y_i} 
\end{equation}

It is worth noting that generally, in I.R.A.,
we can assume:
\begin{equation}
{X_i \over X_j}= {y_i \over y_j}
\end{equation}
namely that the abundance ratios are equivalent to the yield ratios, and this holds also in 
analytical models with infall and/or outflow.
Clearly, eq. (1.17) can be used safely only for elements produced on short timescales, such as $\alpha$-elements, but it fails if applied to elements produced on long timescales such as Fe and N.
These days the  Simple Model is rarely used in describing the chemical evolution of the Milky Way since it does not reproduce the 
G-dwarf metallicity distribution (the G-dwarf problem) as well as the elements produced on long timescales such as Fe (Matteucci, 2001).

\section{Numerical models of chemical evolution}
In the last years a great number of models relaxing I.R.A. 
and the closed box assumption but retaining the constancy of the IMF and the I.M.A., have appeared in the literature.
As an example of the basic equations adopted in such models I show the formulation of Matteucci \& Greggio (1986), which is 
based on the original formulations of Talbot \& Arnett (1971) and Chiosi (1980).

Let $G_i$ be the mass fraction of gas in the form of an element $i$ 
($\sigma_i/ \sigma_{tot}(t_G)$), 
we can write:

\begin{eqnarray}
\dot G_i(t)  =  -\psi(t)X_i(t)  + \int_{M_{L}}^{M_{Bm}}\psi(t-\tau_m)
Q_{mi}X_i(t-\tau_m)\phi(m)dm\nonumber + \\ 
+ A\int_{M_{Bm}}^{M_{BM}}
\phi(m)\cdot[\int_{\mu_{min}}^{0.5}f(\mu)\psi(t-\tau_{m2}) 
Q_{mi}X_i(t-\tau_{m2})d\mu]dm\nonumber + \\ 
+ (1-A)\int_{M_{Bm}}^
{M_{BM}}\psi(t-\tau_{m})Q_{mi}X_i(t-\tau_m)\phi(m)dm\nonumber + \\
+ \int_{M_{BM}}^{M_U}\psi(t-\tau_m)Q_{mi}X_i(t-\tau_m) 
\phi(m)dm+ X_{A_{i}} IR(t) - X_{i} W(t)
\end{eqnarray}
where $A$ is a constant parameter chosen in order to fit the present time SN Ia rate and it lies in the range A=0.05-0.09. The SNIa rate should be based on the existing theories
on SN Ia progenitors which I will summarize in the next section. In the equations above the type Ia SNe are assumed to originate from C-O white dwarfs (WD) in binary systems and $f(\mu)$ represents the distribution of mass ratios in such binary systems ($\mu={M_2 \over (M_1 +M_2)}$) (see Matteucci \& Recchi, 2001 for more details). 
The star formation rate $\psi(t)$ and all the rates in this equations are expressed as functions of the fraction of gas ($G={\sigma_{gas}
\over \sigma_{tot}(t_G)}$).
The quantities $IR(t)$ and $W(t)$ represent the rate of gas accretion and the rate of galactic outflow (wind), respectively, and $X_{A_{i}}$ are the abundances of the accreting material, which
are usually assumed to be primordial (no metals).
Generally, in describing the solar neighbourhood and the Galactic disk one assumes $W(t)=0$, whereas the infall of mostly primordial material is most likely to be responsible for the formation of the disk.

\subsection{Type Ia SN progenitors}

{\bf The single degenerate scenario} is the 
classical scenario originally proposed by  Whelan and Iben (1973), namely C-deflagration in
a C-O WD reaching the Chandrasekhar mass limit, $M_{Ch} \sim 1.44M_{\odot}$,  
after accreting material from a red giant
companion. The progenitors of C-O WDs lie in the range 0.8-8.0$M_{\odot}$, therefore, the most massive binary system of two C-O WDs is the 8$M_{\odot}$   +8
$M_{\odot}$ one. The clock of the system in this scenario is provided by the lifetime of the
secondary star (i.e. the less massive one in the binary system).This implies that the minimum timescale for the appearence of the first type Ia SNe is the lifetime of the most massive  secondary star.
In this case the time is
$t_{SNIa_{min}}$=0.03 Gyr (Greggio and Renzini 1983a; Matteucci \& Greggio, 1986; Matteucci \& Recchi, 2001).\par

{\bf The double degenerate scenario}
consists in the merging of two C-O WDs, due to loss of angular momentum occuring as a consequence of gravitational wave radiation,
which then explode by C-deflagration when the $M_{Ch}$ is reached (Iben
and Tutukov 1984). In this case the minimum timescale for the appearence of type Ia SNe is the lifetime of the most massive secondary star plus the gravitational time delay which depends on the original separation of the two WDs and which is computed according to Landau \& Lifschitz (1962), namely $t_{SNIa_{min}}=0.03 +\Delta t_{grav}$=0.03+ 0.15 Gyr (Tornamb\` e \& Matteucci, 1986). We recall that the gravitational time delay can be as long as several Hubble times.
\par
{\bf The model by Hachisu et al. (1999)} is based on the  
classical scenario of Whelan \& Iben (1973) 
but with a
metallicity effect implying that  no type Ia systems can form 
for [Fe/H]$< -1.0$dex  in the ISM. This implies a much longer time delay for the first type Ia SNe to occur, also because the maximum masses of the secondary stars in the binary systems are assumed to be 
$\le  2.6 M_{\odot}$. Therefore, in this case $t_{SNIa_{min}} = 0.33 $ Gyr + metallicity delay due to the chemical evolution of the considered system. In this scenario, type Ia SNe are not associated with
young stellar populations contrary to what is suggested by a recent search for SNe Ia in starburst galaxies (see Mannucci et al. 2003), which seem to favor the scenario where the progenitors of type Ia SNe can be as high as 8$M_{\odot}$.

Clearly, $t_{SNIa_{min}}$ is a very important parameter for computing galactic chemical evolution since the abundance patterns will depend strongly on this timescale, although the most important timescale is the one for which the type Ia SNe have an impact on the abundances of the ISM.
In succesful models of galactic chemical evolution of the solar vicinity this timescale
is 1.0-1.5 Gyr in the framework of the single degenerate model (Matteucci \& Greggio, 1986). In the other two scenarios this time is longer, especially in the one including the metallicity effect and it varies with different histories of star formation(see Matteucci \& Recchi 2001 and sect. 1.13).

\section{Different approaches to the formation and evolution of the Galaxy}
In the past years several approaches to the calculation of the chemical evolution of the Galaxy were proposed, they are:\par

i) {\bf the serial formation}, where the
halo, thick and thin disk form in a sequence as a continuous process
(e.g. Matteucci  \& Fran\c cois 1989). 
\par
ii) {\bf The parallel formation} where
the various Galactic components start forming at the same time and 
from the same gas but evolve at different rates. This approach
predicts overlapping of stars belonging to the different components
(e.g. Pardi, Ferrini \& Matteucci 1995) but it does not provide a good fit to  the G-dwarf metallicity distribution and the halo star metallicity distribution
simultaneously, as discussed in  Matteucci (2001). 

iii) {\bf The two-infall approach} where
the halo and disk 
form out of two separate infall episodes from extragalactic gas.
Also in this case we predict an  overlap in metallicity between the
different galactic components
(e.g. Chiappini et al. 1997; Chang et al. 1999).
A threshold density ($\sigma_{th}=7 M_{\odot} pc^{-2}$) in the star formation process is also included in the model of Chiappini et al. (1997). 

iv) {\bf The stochastic approach} where
the assumption is made that in the early halo phases, mixing was not efficient 
and pollution from single SNe  would dominate the galactic enrichment
(Tsujimoto et al. 1999; Argast et al. 2000; 
Oey 2000). In this case a large spread in the abundances and abundance ratios at low metallicities is predicted. This predicted spread, however, 
is much larger than observed, especially for $\alpha$-elements.

\section{Observational Constraints}
A good model of chemical evolution should be able to reproduce a minimum number of observational constraints and the number of observational constraints should be larger than the number of free parameters which are: $\tau_H$, $\tau_D$, $k_1$, $k_2$, $\nu$, the IMF slope(s) and the parameter describing the wind, if adopted.

The main observational constraints in the solar vicinity that a good model should repoduce (see Chiappini et al. 2001) are: \par

\begin{itemize}
\item The present time surface gas density:
$\Sigma_G= 13 \pm 3 M_{\odot} pc^{-2}$

\item The present time surface star density $\Sigma_{*}= 48 \pm 9 M_{\odot} pc^{-2}$

\item The present time total surface mass density: $\Sigma_{tot} = 51 \pm 6 M_{\odot} pc^{-2}$

\item The present time SFR:  $\psi_o=2-5 M_{\odot} pc^{-2} Gyr^{-1}$

\item The present time infall rate: $0.3-1.5 M_{\odot} pc^{-2} Gyr^{-1}$

\item The present day mass function (PDMF)

\item The solar abundances, namely the chemical abundances of the ISM at the time of birth of the solar system  4.5 Gyr ago and the present time abundances

\item The observed  [$X_i$/Fe] vs. [Fe/H] relations

\item The  G-dwarf metallicity distribution

\item The  age-metallicity relation
\end{itemize}
And finally, a good model of chemical evolution of the Milky Way should reproduce the
distributions of
abundances, gas and star formation rate along the disk as well as 
the average SNII and Ia rates along the disk (SNII=$1.2 \pm 0.8 \,\, 100yr^{-1}$ and SNIa=$0.3 \pm 0.2 \,\, 100yr^{-1}$)

\section{Time-delay model interpretation}

The difference in the timescales for the occurrence
of SNII and Ia produces a time-delay 
in the Fe production relative to the $\alpha$-elements (Tinsley 1979; Greggio \& Renzini 1983b;
Matteucci 1986). On this basis  we can interpret all the observed abundance ratios plotted as functions of metallicity. In particular, this interpretation is known as time-delay model and can be easily illustrated by Figure 1.3, where we show the predictions of the two-infall model for the chemical evolution of the solar vicinity concerning the [O/Fe]  vs. [Fe/H] relation. We show the 
standard case in which both the contributions to Fe enrichment from type II and Ia are taken into account as well as the cases where only one type of SN at the time is assumed to contribute to Fe enrichment. Both data and models are normalized to the solar abundances (Anders \&
Grevesse 1989). From the Figure 1.3 is evident that only type Ia SNe as Fe
producers would predict a continuous decrease of the [O/Fe] ratio from low to high
metallicities (upper curve), whereas only type II SNe would create a roughly constant [O/Fe] ratio. Therefore, only a model including both contributions in the percentages of 30\% (SNII) and 70\%(SNIa) can reproduce the data.

\begin{figure}
\centering
\includegraphics[width=6cm,angle=0]{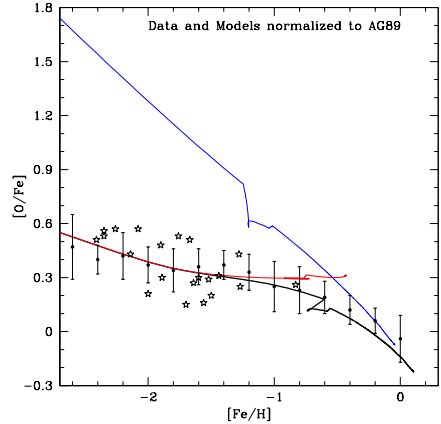}
\caption{Time-delay model.The models and the data are normalized to the solar abundances of Anders \& Grevesse 1989. The thick curve represents the predictions of the standard time-delay model where type Ia SN produce 70\%
of Fe and type II SNe the remaining 30\%. The figure is from Matteucci \& Chiappini, 2003 in preparation)
}
\end{figure}

The time-delay model, with the assumption of an IMF constant in time, can explain the different abundance patterns 
in the halo, disk and  bulge (see Matteucci, 2001).

\section{Common Conclusions from Galaxy Models}

Most of the more recent chemical evolution models agree on several important issues which are:
i) the G-dwarf metallicity distribution
can be reproduced only by assuming that the
formation of the local disk occurred  by infall of extragalactic gas
on a long timescale, of the order of
$\tau_d \sim 6-8$ Gyr
(Chiappini et al. 1997;
Boissier and Prantzos 1999; Chang et al. 1999; Chiappini et al. 2001;
Alib\`es et al. 2001)

\begin{figure}
\centering
\plotone{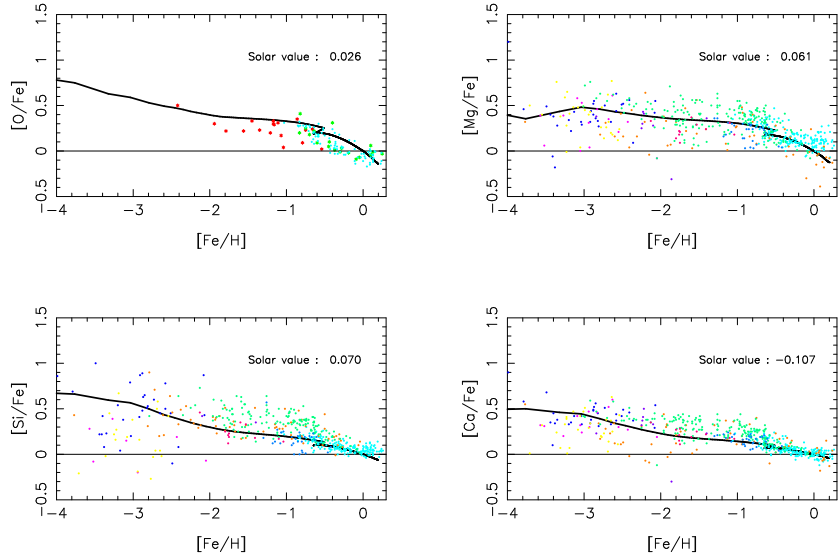}
\caption{Predicted and observed [$\alpha$/Fe] vs. [Fe/H] in the solar neighbourhood.
The model is from Chiappini et al. (1997) whereas the data are from Fran\c cois et al. (2003).
The models are normalized to the predicted solar abundances. The predicted abundance ratios at the time of the Sun formation are shown in each panel and indicate a good fit.}
\end{figure}

ii) The relative abundance ratios [$X_i$/Fe] vs. [Fe/H], interpreted as being due to
time-delay between type Ia and II SNe allow one to reproduce the observed relations
(see Figure 1.4) and to infer the
timescale for the halo-thick
disk formation (corresponding to [Fe/H]=$-$1.0 dex) which should be of the order of
$\tau_h \sim$ 1.5Gyr (Matteucci and Greggio 1986;
Matteucci and Fran\c cois, 1989; 
Chiappini et al. 1997). On the other hand, the external halo formed more slowly, perhaps on timescales of the 
order of 3-4 Gyr (see Matteucci \& Fran\c cois 1992).
In Figure 1.4 we show the predictions of the model developed by Chiappini et al. (1997), 
by adopting the yields of Woosley  \& Weaver (1995) for type II SNe,
with the exception of the Mg yields which have been artificially increased  by a factor of 5 to obtain a good agreement with the solar abundance of Mg, and those of Nomoto et al. (1997) for type Ia SNe (their case W7).The Mg yield in type Ia SNe had also to be increased in order to fit the solar abundances (Fran\c cois et al. 2003, in preparation). The problem of Mg underproduction in nucleosynthesis models is well known and it was pointed out by Thomas et al. (1998).
Figure 1.4 shows clearly that the agreement between the model predictions and data for $\alpha$-elements is quite good and support the time-delay model.

iii) To fit abundance gradients, SFR and gas distributions along the disk, the
disk should have formed inside-out (variable $\tau_D$)
and the SFR should be a strongly varying function of the 
galactocentric distance as in eqs. (1.4), (1.5) and (1.6) (Matteucci \& Fran\c cois 1989; 
Chiappini et al, 1997,2001; Portinari \& Chiosi, 1999,
Goswami \& Prantzos 2000; Alib\' es et al. 2001).

\section{Specific Conclusions from Galaxy Models}
The assumed threshold in the SFR (Chiappini et al. 1997) 
 produces naturally a gap in the star formation process between the end of the halo-thick disk phase and the beginning of the 
thin disk phase. Such a gap lasts 
for $\sim$ 1 Gyr and it seems to be indicated by the observations (Gratton et al. 2000). In particular, 
in Figure 1.5 we show the observed and predicted [Fe/O] vs. [O/H], where it appears that at around [O/H]= $-$0.3 dex there is a lack of stars and then the [Fe/O] ratio rises sharply. This is a clear indication, on the basis of the time-delay model, that there has been a period when only Fe was produced, in other words a halt in the SFR. 
This gap seems to be  observed also in the
[Fe/Mg] vs. [Mg/H] (Furhmann, 1998). However, more data are necessary before drawing firm conclusions on this important point (see also Chiappini \& Matteucci, this conference).

\begin{figure}
\centering
\includegraphics[width=6cm,angle=0]{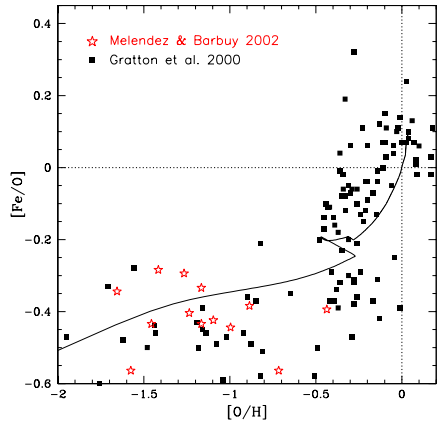}
\caption{Gap in the SFR.The model predictions for [Fe/O] vs. [Fe/H] are obtained by means of the two-infall model (see text). As one can see a gap seems to be evident at [O/H] $\sim -0.3$ dex.}
\end{figure}

\begin{figure}
\centering
\includegraphics[width=6cm,angle=0]{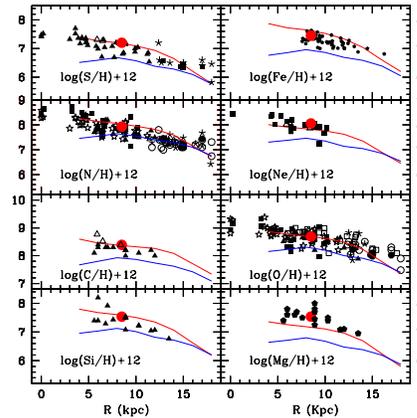}
\caption{Abundance gradients along the Galactic disk and their evolution in time, as predicted by the model of Chiappini et al. (2001). The lower curve of each panel represents the model prediction after 2 Gyr from the beginning of star formation in the thin-disk. The upper curve represent the predicted gradient at the present time.}
\end{figure}

No agreement on the behaviour of gradients in time exists among different authors. In particular, some authors
e.g. (Boissier and Prantzos 1998; Portinari \& Chiosi, 1999, Alib\`es et al. 2001) find a flattening of abundance gradients with time whereas others (Matteucci \& Fran\c cois, 1989; Chiappini et al. 2001) predict  a steepening of the abundance gradients, in agreement with results from chemo-dynamical models (Samland et al. 1997). 
The difference between the two different approaches assumed by the various authors could  perhaps reside in the different star formation and/or infall laws adopted for the Galactic disk (see Tosi, 2000 for a discussion of this point). Data from planetary nebulae (PNe) of different ages can help in solving this problem; recently Maciel et al. (2002) suggested a flattening of the gradients with time. In any case, all the models
predict a very small evolution in the last 5 Gyr.

\section{Star formation rate in galaxies}
It is very important to understand that different histories of star formation determine
the abundance patterns in galaxies. In particular, the [$X_i$/Fe] vs. [Fe/H] relations are strongly influenced by the SFR  which
influences the temporal growth of the Fe abundance.

In other words, the time-delay model coupled with different SF histories
implies different timescales for the bulk of Fe production from type Ia SNe,
thus producing
different [$\alpha$/Fe] vs.
[Fe/H] relations in different objects.
The typical timescale for type Ia SN enrichment can be defined
as the time when the SN Ia rate reaches the maximum
$t_{SNIa}$ (Matteucci \& Recchi 2001).
This timescale depends upon the progenitor lifetimes, the IMF and the SFR.
For an elliptical galaxy or a bulge of a spiral
with high SFR the maximum in the type Ia SN rate is reached at
$t_{SNIa}= 0.3-0.5$ Gyr, whereas 
for a spiral like the Milky Way a first maximum in the SNIa rate is reached at  $t_{SNIa}=1.5$ Gyr and a second maximum, if one uses the two-infall model, at 
$t_{SNIa}=4-5$ Gyr. Therefore, for the Milky Way 1-1.5 Gyr is the time at which SNe Ia are no more negligible in the process of chemical enrichment and it corresponds to the change in slope observed in the [el/Fe] vs. [Fe/H] relations (Figure 1.4) and to the end
of the halo phase.
For an irregular galaxy, where the SFR is assumed to proceed more slowly than
in the solar vicinity the timescale for SN Ia enrichment is $t_{SNIa}=7-8$ Gyr.
Therefore, it is worth repeating that the $t_{SNIa}$ is different in different galaxies, since often in the literature it is adopted a universal timescale of 1 Gyr!
On the basis of that we expect that
objects where the SFR proceeds very fast such as in the spheroids (bulges and ellipticals), the [$\alpha$/Fe] ratio stays flat for a larger 
metallicity interval than in systems with slower star formation such as the Milky Way, and that eventually the [$\alpha$/Fe] ratios in irregulars, where the star formation rate has been less efficient than in spiral, decreases almost continuously, as shown in Figure 1.7. The models shown in Figure 1.7
contain the same nucleosynthesis prescriptions and differ for the star formation history.

\begin{figure}
\centering
\includegraphics[width=6cm,angle=0]{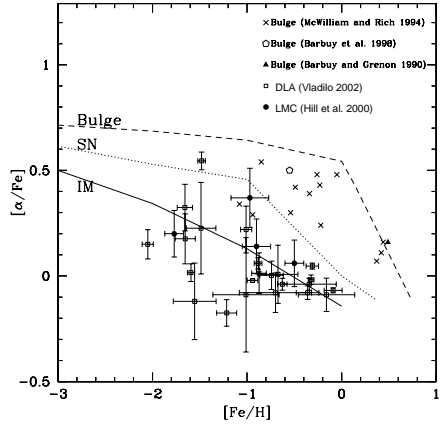}
\caption{The  predicted [$\alpha$/Fe] vs.
[Fe/H] in different objects. Data for the Galactic bulge, the LMC and Damped Lyman-$\alpha$ systems are shown for comparison.}
\end{figure}

\section{High-redshift objects}
In Figure 1.7 we show also some observational data relative to the Galactic bulge, the LMC and some Damped Lyman-$\alpha$ systems (DLA). As one can see, our predictions seem to reproduce the data for the Galactic bulge and for LMC and also suggest that DLAs could be the progenitors of the present day irregular galaxies, since most of DLAs, once their abundances are corrected for the effect of dust, show low [$\alpha$/Fe] ratios
at low metallicities (Vladilo, 2002; Calura et al. 2003). On the other hand, Lyman-break galaxies such as cB58 
(Pettini et al. 2002) show an abundance pattern compatible with a young spheroid, either a galactic bulge or a small elliptical, suffering a galactic outflow,
as recently shown by Matteucci \& Pipino  (2002).
Therefore, the [$X_i$/Fe] vs. [Fe/H]  diagram represents a very  useful tool
to infer the nature of high redshift objects of which we know just the abundances and abundance  ratios.

\end{document}